# Can rhythm be touched? An evaluation of rhythmic sketch performance with augmented multimodal feedback

Feng Feng[a,], Tony Stockman [a], Shang Kai [b]

[a] *School of Electronic Engineering and Computer Science, Queen Mary University of London* [b]*School of Design, Shandong Jiaotong University*

**Abstract**

Although it has been shown that augmented multimodal feedback has a facilitatory effect on motor performance for motor learning and music training, the functionality of haptic feedback combined with other modalities in rhythmic movement tasks has rarely been explored and analyzed. In this paper, we evaluate the functionality of visual-haptic feedback in a rhythmic sketch task by comparing it with other multimodal feedback and a baseline condition on two interfaces. Results showed an equivalent facilitatory effect between the VH condition and other multimodal conditions. Further, we examine the possibility of accessing the quality of task execution through kinematic analysis. Based on participants' speed profiles, we investigated the quality of motor control and movement smoothness under different feedback conditions. Results revealed better motor control ability with auditory feedback and improved movement smoothness with haptic feedback. Finally, we propose that haptic feedback can be integrated with other modal stimuli for different interaction purposes, and that kinematic analysis can be a complementary approach to gesture analysis as well as providing subjective evaluation of interaction performance.

# 1. Introduction

Augmented visual-audio feedback has many advantages for rhythm perception and has been shown to have facilitated motor performance in situations such as sports training, rehabilitation and motor skill refinement [1, 2, 3].

However, auditory feedback can be masked by environmental noise in sports training or in other noisy environments [4] and can be intrusive to non-stakeholders in shared spaces and causes awkwardness to end users [2]. Tactile feedback, in comparison, can overcome this disadvantage by representing effective feedback in a private and inconspicuous manner [5]. However, the question of whether visual-tactile feedback is able to provide equivalent facilitation for rhythm perception and motor control is unknown. The current research aim to address this question by evaluating motor performance under augmented visual-tactile feedback in a rhythmic sketch task. In addition, we are also interested in analyzing the motor performance under different scenarios, which afford two scales of motion range. The first one supports relatively small and fast hand movements, with the focus on possible scenarios involving fine motor function, such as bimanual coordination practise or music performance [6, 7]. The second one supports relatively large and slow arm movement, with the focus on possible scenarios involving gross motor function, such as upper extremity rehabilitation [8, 9].

To examine the effect of visual-tactile feedback on rhythmic motor control, standard evaluation techniques from human-computer interaction (HCI) are adopted, which include measurements of time to complete task, the accuracy of task fulfilment, as well as subjective evaluation [10, 11, 7]. However, to examine the quality of movements under different augmented feedback conditions, continuous time series data of motor performance is required. Therefore, a rhythmic sketch task, rather than rhythmic tapping, [12] or clicking task [7] was adopted. Based on this data set, we conducted kinematic analyses as a complementary approach to task performance evaluation as well as usability evaluation. The purpose is to compare the evaluation approach based on time-specific output, e.g. gestures in rhythm, with the one on less time-specific output, e.g. subjective user experience [13].

The first contribution of the present paper, to our knowledge, is that for the first time, we have evaluated the effect of augmented visual-tactile feedback on a rhythmic sketch task with observations on both the fine motor skill, e.g. hand movement, and the gross motor skill, e.g. the arm movement.

Secondly, the present paper adopted kinematic analysis as a different approach for task performance and usability evaluation, which in this case, complimented the conventional subjective usability evaluation with a detailed analysis of the rhythmic gestures. More broadly, this approach also sheds light on the design of augmented kinematic features for interactive tasks, which may contribute to improving and maintaining motivation for motor skill practise [14, 15] and/or home-based rehabilitation [2, 9].

The paper starts by introducing the research background, including the use of multimodal information as a feedback strategy in current motor learning and interaction contexts, as well as the neuroscience foundations of rhythm perception and motor execution. In the subsequent sections, the research aim, scope and the evaluation methods are described, followed by the details of the experimental methods, results and discussions for each of the two research studies. Finally, a general discussion of the augmented multimodal feedback and the corresponding usability evaluation method are presented.

## 2. Background

### 2.1. Augmented feedback for sensory-motor performance

Augmented feedback, also refered to as extrinsic feedback, is defined as in-formation feedback that cannot be elaborated without an external source being used for the stimuli [16]. It is one of the instructional strategies that makes use of elements of gestural or bodily movement, such as trajectory, speed, force exception etc., to increase self-awareness of motion quality and facilitate motioncorrection in the context of music performance and motor skill learning [1, 17].

Augmented feedback can be classified according to the time at which the feedback is given, The specific movement features amplified by the concurrent feedback and/or the task being executed [18]. Previous research has shown that simple motor tasks which have one degree of freedom and involve smaller movements facilitate performance during skill acquisition, but the effect dissipates the moment the feedback is withdrawn. This phenomenon is explained by the guidance hypothesis [18, 19, 20]. However, for complex motor tasks, which have several degrees of freedom and involve larger movements, concurrent visual or auditory feedback has been shown to reduce cognitive demand during task execution, and is effective in improving complex motor performance when the feedback is withdraw [18, 21, 22]. In comparison, terminal feedback provides information at the end of or after task execution[16, 1], which in some cases has been shown to have a better facilitatory effect on simple motor task acquisition rather than has been seen with complex motor tasks [23, 24].

Another way to classify augmented feedback is based on the number of modalities involved in feedback representation, such as visual, auditory and/or visual-haptic feedback. Previous research demonstrated that modality specificity influences motor performance, since sensory modalities have specific sensibility for processing different forms of information. For example, in the acquisition of spatial information, visual perception dominates other sensory modalities if it has not been degraded or masked. Thus the augmented visual feedback strategy has been the primary consideration (most explored approach) for tasks requiring high spatial accuracy [1]. However, the visual modality has a relatively low temporal resolution compared with auditory and haptic senses [25].

In tasks involving rhythmic finger tapping, visual modality showed a higher synchronisation threshold than auditory modality, which correlated with an inferior motor performance on rhythm entrainment [26]. Auditory perception is very precise in capturing temporal information, especially for periodic, regulatory and speeded events [27]. It has been shown that in tasks containing periodic or rhythmic audio-visual information, auditory perception modulates or can even bias visual processing of relevant information [28]. The haptic modality, on the other hand, showed comparable sensitivity to both temporal and spatial information, with a particular effect of regulating periodic movement in the context of motor training [4, 29] as well as music performance [17]. However, more systematic evaluation of the effectiveness of haptic feedback, especially vibrotactile feedback, is required in the context of motor performance.

In recent years, the effectiveness of augmented multimodal feedback on motor performance has gained interest [30, 31, 15]. The rationale lies in the behavioural and neuroscience research outcomes. It is well established that multimodal perception has advantages over unimodal perception in terms of better precision and faster processing speed [32]. Design research studies implemented multimodal approach in many applied fields such as driving security [33], military training [34], surgery simulation [35], and sports training [36]. These studies demonstrated that well designed multimodal feedback supports faster cognitive response and more accurate judgement in goal-oriented interactive tasks. These observations can be interpreted in the light of multiple-resource theory, which states that distributing information across separate modalities reduces cognitive load and thus broadens the bandwidth available for information processing [37]. Indeed, many research studies have shown that as tasks become more complex, multimodal representations are increasingly welcomed by users [38]. In research more relevant to motor performance, an operation simulation with augmented visuohaptic feedback prevailed over visual feedback in a needle insertion task [39]. In an interactive game that involved throwing a virtual ball towards a goal, visuohaptic feedback combined with sonification significantly improved interaction accuracy as well as perceived intuitiveness of the interface [14]. However, to our knowledge, systematic evaluation of the effect of augmented multimodal feedback on periodic sensory-motor performance has received little attention in previous literature.

*2.2. Neuroscience and behavioural background on rhythm perception*

Research on neural processing on temporal regularities has provided evidence that the functionality of rhythm perception and motor timing control share the same area of the brain [40], which is responsible for predicting sequential patterns of sensory inputs in time [41], as well as for preparing forthcoming sequential movements [42].

Bengtsson et al. demonstrated that auditory perception of rhythms induced more activation in relevant motor areas of the brain [40]. In the experiment, three types of rhythmic sequence and one unpredictable sequence were presented to participants while their brain activity was recorded through functional magnetic resonance imaging (fMRI). The isochronous sequence was composed of an audio sequence that has equivalent intervals duration between notes. The metric sequence has varied interval durations with an integer ratio. For example, a sequence has the notes onset at 0ms, 600ms, 800ms, 1600ms and 2000ms sequentially, and the interval ratio of this sequence would be 3:1:4:2. The non-metric sequence, in comparison, has decimal ratio interval durations. Take the same case as an example, a non-metric sequence could have ratio intervals of 3:1.5:4:2.

Lastly, the unpredictable sequence was composed of notes that have randomly distributed intervals. Results of this experiment showed that the motor area in the brain was activated by all three rhythmic sequences but not by the unpredictable sequence. This observation and together with more recent supportive research [43, 44] confirms that auditory perception of rhythmic patterns could indeed modulate motor responses.

There is also evidence that visually perceived sequential events could also activate motor brain areas [45, 41, 42]. Penhune et al. used positron emission tomography (PET) to test whether both the auditory and visual sequential stimuli can activate motor areas with a similar pattern [45]. With a rhythm re-producing task performed by finger tapping, results of both the behavioral performance and the PET images confirmed a similar activation pattern of timed motor responses [45]. In another experiment, Schubots and Cramon used sequences of visual stimuli that varied in size to investigate the mechanism between visual (as well as auditory) pattern prediction and premotor activation. The fMRI result reflected that the prediction of the visually perceived pattern activated the ventrolateral premotor cortex, which is commonly associated with object grasping and manipulation [41].

At the behavioural level, it has been repeatedly demonstrated that the auditory system is more sensitive to rhythm perception than the visual system on both the non-isochronous and isochronous rhythms [25, 46, 47, 48]. However, the type of rhythm could influence the sensibility of visual modality on rhythmic prediction. For instance, a rhythm in beat-based structure, in which stimulus onsets are aligned with beat locations, has been shown to better support visual prediction than sequences that do not follow that structure [49]. A more recent study showed that people's expertise has an effect on the extent of their auditory advantage on rhythmic motion perception [50]. In the experiment, participants were asked to listen to the audio or watch a video of tap dance, e.g. a sequence of footsteps, and detect whether the rhythm was regular or irregular. Only expert participants produced a statistically significant higher rate of correct answers in the audio condition than in the video (visual) condition. In addition to the auditory and visual modalites, the effect of tactile rhythmic cues on rhythmic motor function have rarely been systematically investigated on both the neurophysiological and the behavioural levels, other than a few attempted applications in the field of gait rehabilitation [51, 52], which showed inconsistent results of the effect of tactile cueing on gait performance.

## 3. Research aims, scope and evaluation context

### 3.1. Research aims

The main objective of the current research is to evaluate whether, combined with visual stimuli, tactile feedback has an equivalent facilitatory effect as auditory feedback does on rhythm perception and rhythmic movement; and whether similar behavioral patterns can be observed on different scales of motion on two interfaces. The second objective is to examine whether kinematic analysis can be a complementary method for usability evaluation, based on time series data of continuous movement.

It is commonly acknowledged that haptic modality is the only sensory modality that can act on the external world to make changes, while perceiving those changes simultaneously [53]. Thus the execution of an action and the tactile perception in real-time should be coordinated in a way that optimizes the spatial-temporal precision of a movement, especially when the movement follows a periodic and rhythmic pattern. We hypothesized that in the presence of visual information, tactile feedback could also play an important role in enhancing rhythm perception [54], as well as rhythmic movement control [7]. Therefore, by manipulating combinations of modal feedback, we can explore whether or not, and to what extend rhythmic motor performance varies accordingly.

### 3.2. Scope of investigation

Previous research on sensory-motor synchronization frequently adopted a finger-tapping task which were evaluated based on discrete time series data [17, 12]. However, since we wanted to perform motion analysis in relation to augmented multimodal stimuli, continuous time series data is required. Thus we adopted a rhythmic sketch task. Two user studies were conducted with this task on two types of interfaces, which respectively support a fine motor skill, e.g. rhythmic sketch by hand, and a gross motor skill, e.g. rhythmic sketch by arm. Study 1 was conducted based on a sketchpad interface, which affords fine movements, and study 2 was conducted using a tangible handle combined with a wall-display, which affords gross movements.

Empirical studies have suggested that people are more able to entrain and reproduce metric rhythms rather than non-metric rhythms [55, 56], thus metric rhythms are viewed as being easier to learn than non-metric rhythms. In order to minimize the confounding factor of complexity among non-metric rhythms, we adopted only metric rhythms for the motor tasks. Details of how this confound was handled are given later in the discussion of rhythm in subsection 4.3

*3.3. The rationale for evaluation methods*

Standard HCI evaluation methods involve a rigorous and structured analysis based on measurements such as reaction time, efficiency, accuracy and correction rate, which are measurements that focus on the task performance analysis of interfaces and on interaction outcomes that can only be revealed at the post-interaction phase. However, with a specific focus on understanding what happens during the execution of an action, such as sketching or gestural input, evaluation at the end of task execution reflects only partial aspects of the interaction behaviour. The other aspects, for instance, what is the quality of an input gesture, how smoothely is the gesture performed, what is the users intention based on the current interaction state, what is his/her level of self-awareness and confidence, an action cannot be fully unveiled and explained with these post-interaction measures. Some light on these issues can be provided through the use of post-interaction questionnaires or interviews, but these approaches are subject to issues of self reporting and lack objectivity. To address this problem, we employed kinematic analysis, which is commonly used in the fields of biomechanics and motor training, for understanding characteristics of motion [57, 1, 15]. The features of movement that are commonly evaluated during kinematic analysis include position [1], velocity and acceleration [58], force exertion [15] etc. In this study, we adopted a similar approach to analyse the usability of augmented multimodal feedback with a rhythmic sketch task, and compare the results with subjective usability evaluation.

## 4. Methods

Since the two studies reported here employed identical experimental methods, in this section details of the experiment design will be described for both studies.

The experiments used a within-subject design, with the augmented terminal feedback in different modality combinations as the manipulating factor (independent variable). The four experimental conditions are: VHA condition, which presents terminal feedback through visual, haptic and auditory modes simultaneously; the VH condition, which combines visual and haptic feedback, and the VA condition, which combines visual with auditory feedback. The fourth condition was employed as a baseline condition, which provides only the visual feedback, for the purpose of isolating the visual dominance effect from the manipulation effect [59, 60]. If the results from the above three manipulation conditions show a same correlation pattern, through comparison with the baseline condition, we can deduce whether the similarity is the result of manipulation or of visual dominance effect.

For the purpose of minimizing learning effects, the four types of augmented feedback with four rhythms was randomized by a 4*4 Latin square, resulting in sixteen experimental trials. Each participant did the sixteen trials in three experimental blocks, which was randomized again by a 4*3 Latin square. In total, 48 trials were completed by each participant.

### 4.1. Participants

Thirty-two volunteers (sixteen females, sixteen males), aged from 18 to 25, participated in the two studies. Eight females and eight males joined study 1, and the other sixteen participants joined study 2. They were recruited through the university's email lists and social network application. According to the demographic questionnaire, all participants have no visual or auditory disorder, no severe bone or joint injury on their upper limbs in the last three month. They all read and signed a consent form with an ethical approval attached, which was authorized by (blinded for review) Ethics Research Committee.

*4.2. Experimental task and procedure*

The task was to learn rhythmic patterns that were presented once with augmented terminal feedback, and then to reproduce the rhythm by performing rhythmic sketching using sketchpad for study 1 figure 1, or using the tangible handle on the wall-display for study two figure 2. The augmented feedback provided during the learning phase was displayed again during the reproduction phase that was followed.

*4.3. Stimuli*

There were four metric rhythms in the rhythmic sketch task. All the rhythms were composed of five identical musical notes which have different interval ratios and without any accent in the sequence. Rhythm 1 was composed of the interval ratio of 6, 1, 3, 3, rhythm 2 was composed of 3, 6, 1, 3, and the interval ratio of 3, 3, 6, 1, as well as 1, 6, 3, 3 was assigned to rhythm 3 and 4 respectively (Figure 3). The Onset duration of each of the five notes was 250ms, between the notes were the interval ratio times 100ms. Each of the four metric rhythms had a total duration of 2550ms.

The visual rhythmic patterns were displayed as two horizontal circles that flashed one after another consecutively. Since the rhythmic patterns composed of five signals, the visual flash pattern always followed a left-right-left-right-left order (Figure: 4). The haptic rhythm was presented through a Pico Vibe motor embedded in an armband but have no direct contact with participants skin.

The auditory rhythm was presented through a Mac pro speaker with output volume at level 12. For the multimodal feedback conditions, the signal onset of all output channels was calibrated to the same time.

*4.4. Apparatus*

The figure 1 and figure 2 shows the experimental apparatus for study 1 and study 2. In study one, the visual rhythms were presented through a Mac pro screen, with the angle adjusted to 135°. The Apple magic trackpad 2 was used as the sketchpad during the experiment. Participants were encouraged to move their hand and finger without arm movements by making sure their wrist touch the table. In study 2, visual part was projected on the wall as a wall-display. The distance between participants and the wall-display was 4 meters, with a camera set in the middle with a distance of 1.2 meters to participants. The area of projection is 1500 × 1310mm$^2$. The rhythmic sketches in study 2 were performed by drawing in the air by holding a tangible object figure 2, and the trace of the sketch were projected on the wall-display in real-time.

*4.5. Data acquisition and processing*

Coordinates and time stamps of hand and arm movement data were recorded during the experiment at the rate of 125 per second. Two sets of information were extracted from the recording. First, the temporal information concerning the reproduced rhythm was calculated based on the changing direction of hand movements. This information was used for measuring the correlation between the sample rhythm and the reproduced rhythm. Second, the speed of hand and arm movement during the rhythmic sketch task was extracted by taking the first order derivative of the movement trajectory. This measurement was used for kinematic analysis of the continuous hand and arm movement. Based on this calculation, two more measures were derived: The first is dynamic time warping (DTW) distance [8, 15], which is an indicator of the level of closeness or similarity of the movements, and the other is the number of velocity peaks, which is an established indicator of movement smoothness [10, 58].

**5. Results**

### 5.1. Task performance results

#### 5.1.1. Pearson's r value

The detailed statistical results from the Pearson correlation coefficient test with $\alpha$ = 0.05 can be seen in table 1 and figure 6. The higher the correlation coefficient values, the better the temporal accuracy of the reproduced rhythm through sketching. For the hand movement on the sketchpad in study 1, the VA, VH and VHA conditions induced more accurate rhythmic movement than the V condition in three out of four rhythms as shown in table 1. For the arm movement with the tangible handle on the wall-display in study 2, the same results were obtained from the three manipulation conditions. However, the manipulation effect was less obvious in this case.

These results reflect that first, the improved motor performance was indeed due to the VH feedback and VA, VHA feedback, and secondly, the temporal accuracy obtained for the rhythmic movement of the arm over a relatively large

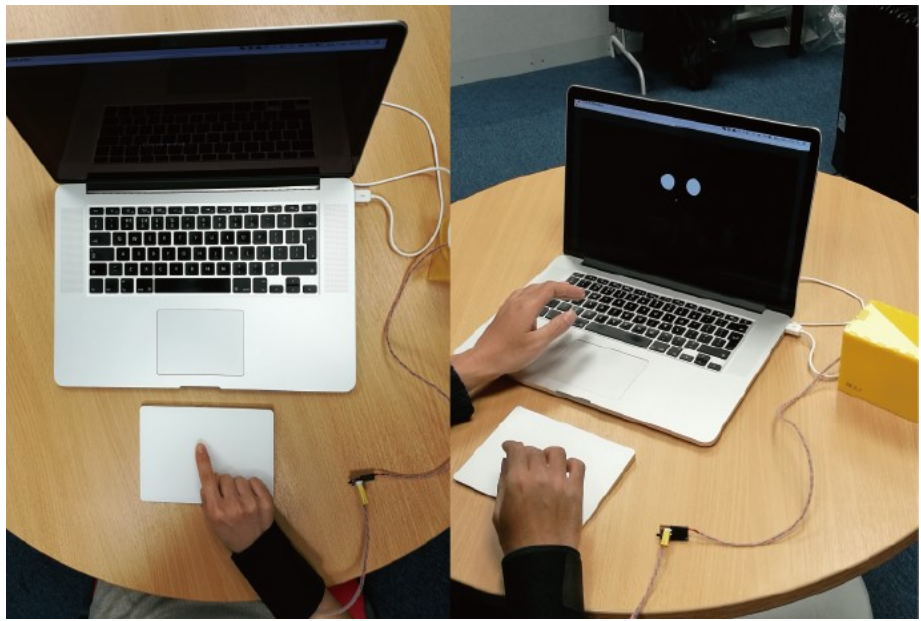

Figure 1: The experimental apparatus for the hand-based task, which composed of a screen, a sketch pad and a wrist band with a motor embedded.

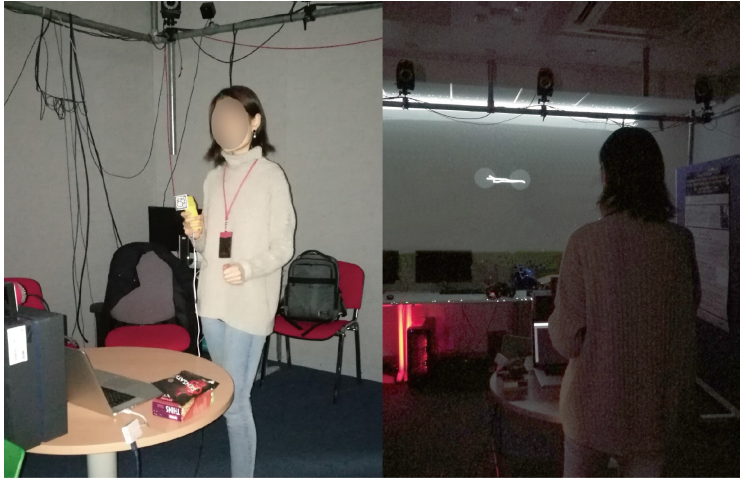

Figure 2: The experimental apparatus for the arm-based task, which composed of a projector, an infrared camera and a hand-held device.

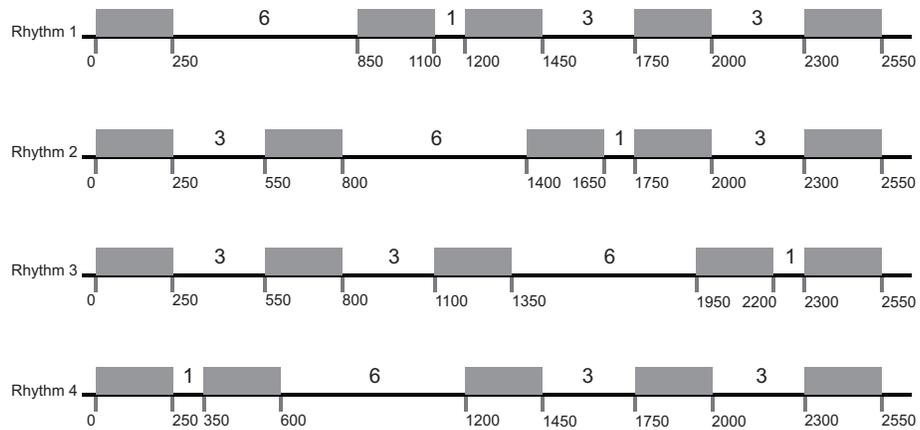

Figure 3: The four metric rhythms used in the rhythmic sketch task. The black lines indicate total duration of the the rhythm. Each of the gray squares indicate the stimuli onset duration, which are 250ms.

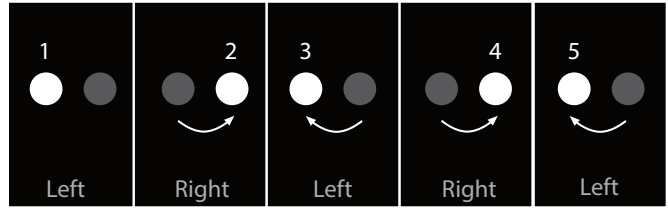

Figure 4: The visual flash pattern in experiment trials.

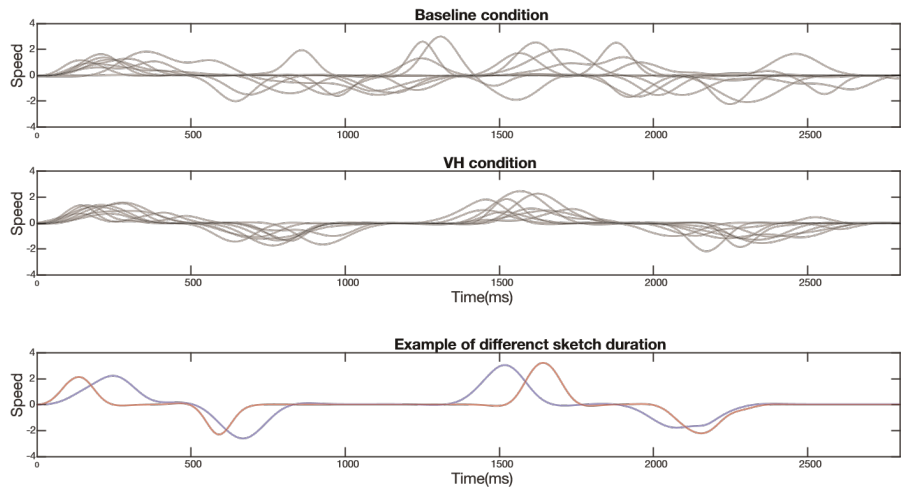

Figure 5: The speed profiles produced during the interaction task. The top figure shows the speed profiles produced under the baseline V condition. The middle one shows the speed profiles with the VH condition. The bottom figure shows two speed profiles that performed with different tempo. The red line has slower tempo than the blue line.

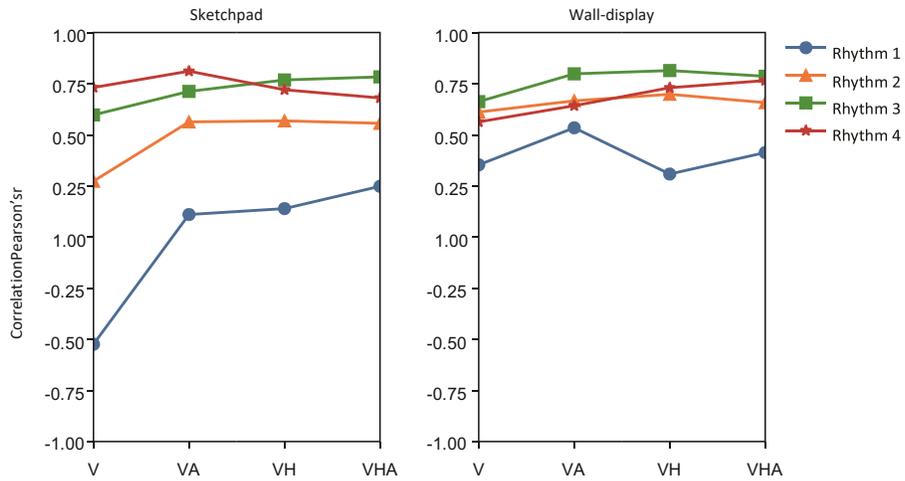

Figure 6: The results of the Pearson correlation coefficient test. The left figure shows the results of the hand-based task on sketchpad, and the left one shows the arm-based task on the wall-display.

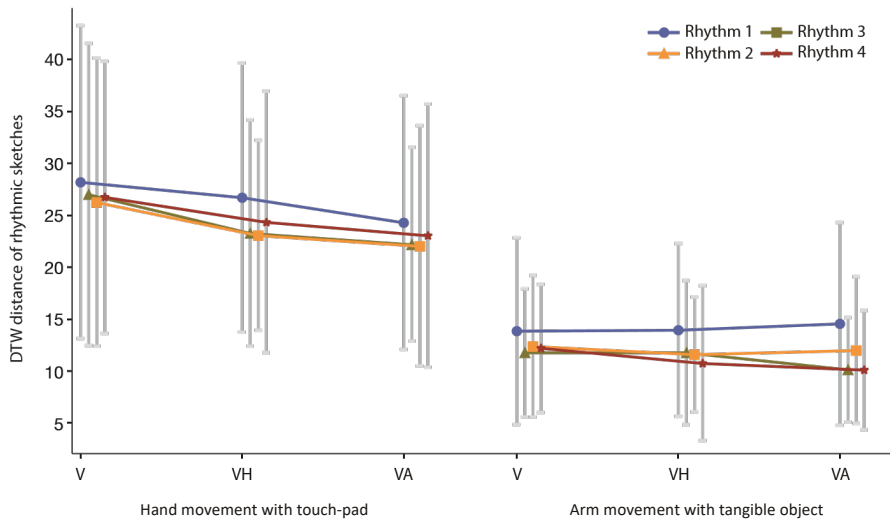

Figure 7: The dynamic time warping (DTW) distance on rhymes sketches between V, V, VA conditions and the VHA condition. The left side shows the results produced from hand-based tasks and the right side are from arm-based tasks. The error bars are standard deviations.

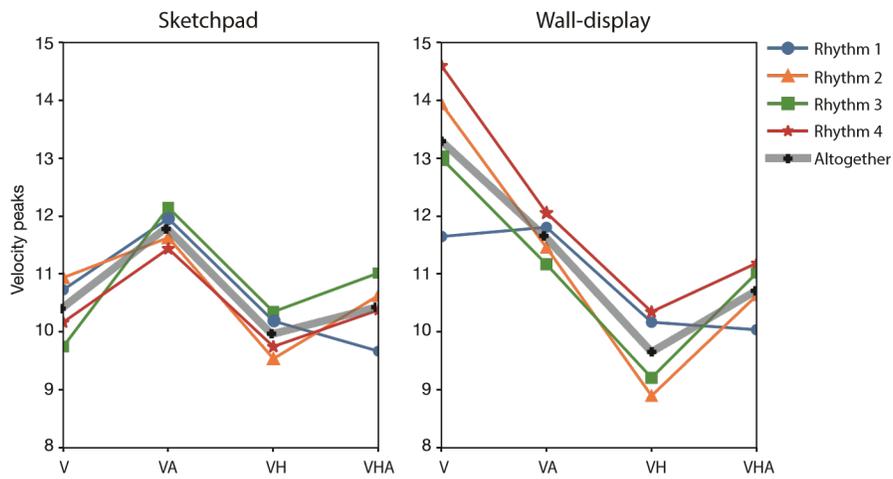

Figure 8: The number of velocity peaks produced in hand-based tasks with the sketchpad (left), and in the arm-based tasks with the wall-display (right). The blue line represents the result under the rhythms 1, orange line represents the rhythm 2, the green line represents the rhythm 3 and the red line represents the rhythm 4. The black bold line shows the averaged velocity peaks of all four rhythms.

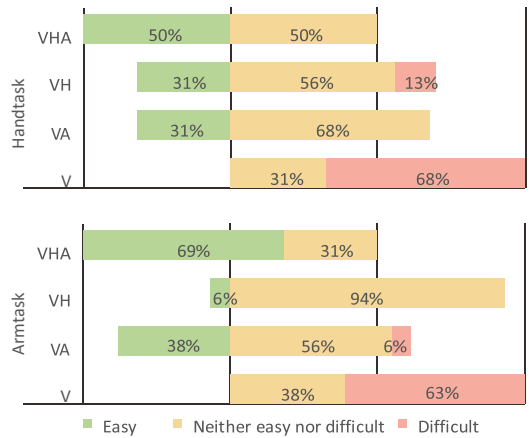

Figure 9: The frequency of the rating in terms of the subjective evaluation on task difficulty.

Table 1: Correlations between sample rhythm and reproduced rhythm

| | **Touch-pad** | **Wall-display** |
| --- | --- | --- |

| | Feedback | Pearson's r | P value | Pearson's r | P value |
|---|---|---|---|---|---|
| Rhythm 1 | | -0.524 | 0.021 | 0.353 | 0.001 |
| | V | 0.11 | 0.072 | 0.534 | 0.001 |
| | VA | | | 0.308 | 0.006 |
| | VH | 0.139 | 0.027 | | |
| | VHA | 0.248 | 0.005 | 0.413 | 0.068 |
| Rhythm 2 | V | 0.273 | 0.631 | 0.611 | 0.002 |
| | VA | 0.563 | 0.01 | 0.666 | 0.002 |
| | VH | 0.568 | 0.043 | 0.698 | 0.00 |
| | VHA | 0.556 | 0.002 | 0.657 | 0.002 |
| Rhythm 3 | V | 0.597 | 0.09 | 0.662 | 0.005 |
| | VA | 0.712 | 0.00 | 0.798 | 0.00 |
| | VH | 0.768 | 0.00 | 0.814 | 0.00 |
| | VHA | 0.783 | 0.00 | 0.786 | 0.00 |
| Rhythm 4 | V | 0.731 | 0.008 | 0.563 | 0.008 |
| | VA | 0.811 | 0.00 | 0.642 | 0.009 |
| | VH | 0.72 | 0.013 | 0.73 | 0.00 |
| | VHA | 0.68 | 0.005 | 0.765 | 0.00 |

Correlation is significant at .05 level

motion range were better than those obtaine for the small-scale hand move-325
ments.

*5.2. Kinematic results*

The kinematic analysis is based on participants' speed profiles as shown in figure 5 (top two charts). Since similar movement patterns may vary in durations, in other words, the reproduced rhythms may have consistent note

330    onset intervals but with either slower or faster tempo, as shown in figure 5 (bottom chart). Direct comparison based on one single time duration cannot reflect the accuracy of rhythms reproduced under different feedback conditions. DTW is one of the ways to compare the similarity between two sequences of time series data which have different lengths. In this case, the time series data 335 are the movement speed profiles produced with different augmented feedback. By calculating the warping distance of the speed profiles, we could understand the similarities of movements executed during the interaction.

Table 2: DTW distance of V, VH, VA to VHA

|  |  | **Touch-pad** | | | **Wall-display** | | |
| --- | --- | --- | --- | --- | --- | --- | --- |
|  |  | V | VH | VA | V | VH | VA |
| Rhythm 1 | Mean | 28.81 | 27.33 | 24.91 | 13.86 | 13.95 | 14.56 |
|  | SD | 16.42 | 14.13 | 13.21 | 8.95 | 8.31 | 9.80 |
| Rhythm 2 | Mean | 27.00 | 23.26 | 22.17 | 11.77 | 11.77 | 10.13 |
|  | SD | 14.42 | 10.86 | 9.23 | 6.14 | 6.90 | 5.02 |
| Rhythm 3 | Mean | 26.24 | 23.05 | 22.02 | 12.38 | 11.59 | 11.99 |
|  | SD | 13.70 | 9.16 | 11.50 | 6.73 | 5.46 | 7.02 |
| Rhythm 4 | Mean | 26.74 | 24.33 | 23.04 | 12.21 | 10.74 | 10.11 |
|  | SD | 12.98 | 12.51 | 12.67 | 6.17 | 7.45 | 5.72 |

*5.2.1. DTW distance*

The DTW technique was used to map V with VHA, VH with VHA, and VA with VHA respectively for each of the four rhythms. The smaller the warping distance between two time series of two rhythmic sketches, the better the consistency and higher similarity between the two corresponding speed profiles. Table 2 lists numeric values of DTW distance with different rhythms. First, it shows that for hand movement with the touch-pad, VA to VHA warping had the least distance values across all four rhythms. Second, for the arm movement with the tangible handle on the wall-display, warping distances between feedback conditions and rhythms were close. Third, in general, the larger scale rhythmic sketches made by arm movements produced smaller warping distances

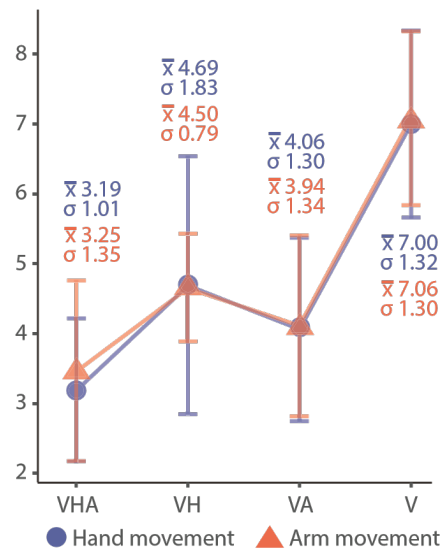

Figure 10: The score of the subjective evaluation on task difficulty. Blue colour represents the evaluation of hand-based tasks, and the red colour represents the arm-based tasks.

than the sketches made by hand movements as shown in figure 7.

### 5.2.2. Movement smoothness

The number of velocity peaks is an indicator of movement smoothness. A perfect rhythmic sketch for the current five-beats rhythms should have four velocity peaks correspond to four movement phases, while the velocity remain zero at the left and right ends as shown in figure 4. However, human motion control is not generally so accurate as to produce such a movement profile. Performing an action composed of many small and large sub-movements has the effect of contributing to an increased number of velocity peaks [10, 58]. In this particular case, small shifts, tilting or jerky movements may happen at the two ends where the velocity tend to be zero. Those sub-movements outside the four main movement phases thus should not be included. One way to filter out the sub-movements outside of the main phases is to find the velocity threshold that is larger than the largest sub-movement and smaller than the least main movement phase. The mean velocities in all four rhythms are in this range and thus were used to remove the sub-movements that fall outside of the main movement phases. The results can be seen in figure 8 and the detail information is listed in the table 3.

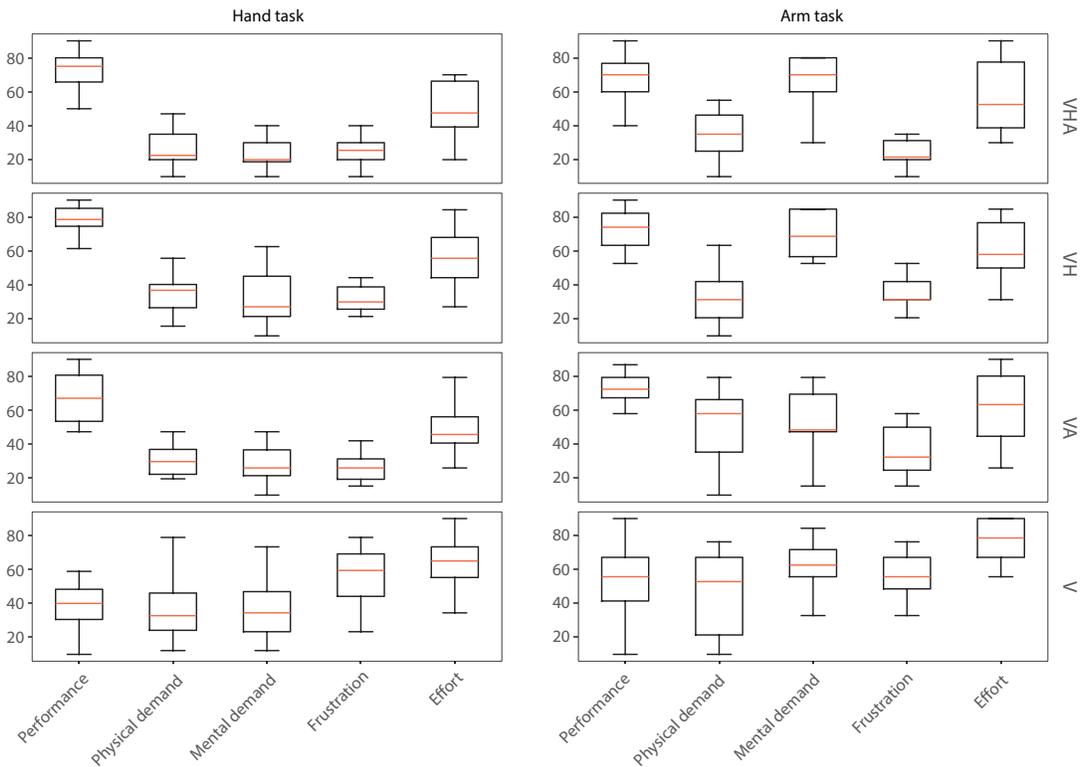

Figure 11: NASA-TLX assessment results. The left side figure shows the hand-based tasks and right side figure shows the arm-based tasks.

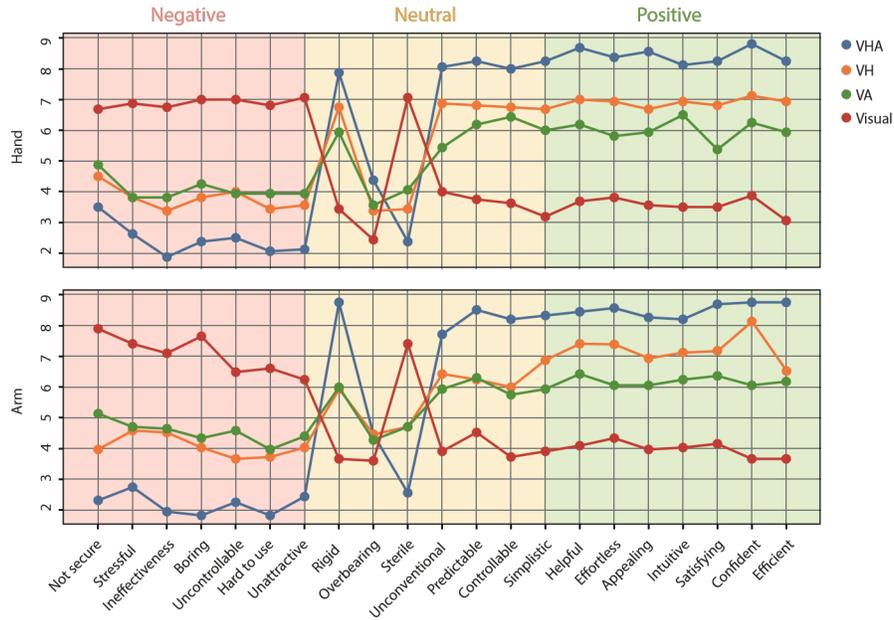

Figure 12: The user experience evaluation on four augmented feedback based on the Reaction card.

### 5.3. Subjective evaluation results

#### 5.3.1. Usability results

We collected participants' subjective evaluations of task difficulty through post-task questionnaires. They were asked to consider their rating of ease-ofuse based on their performance with four types of augmented feedback. The rating range was from very easy (scored as 1) to very difficult (scored as 9). For both the hand and arm interactions with the two interfaces, the ratings of the difficulty for the VHA condition was lowest, followed by the VA condition, VH and base-line conditions V. Detailed information can be seen in figure 10. A Friedman test showed a statistically significant difference in ratings across the four feedback conditions for hand movement: $\chi^2(3, n = 16) = 32.02, p = .000$, and for arm movement: $\chi^2(3, n = 16) = 32.73, p = .000$. Post hoc comparisons indicated that for both the hand and arm movement trials, the score for the VHA condition was significantly lower than the VH, VA and V conditions; the VH and VA conditions were also significantly lower than the V condition. The statistical results can be viewed in table 4.

To get an insight of how the ratings varied across the different types of feedback, the frequency of the rating split into three levels were calculated and are shown in figure 9. Ratings from 1-3 were sorted as easy, and from 4-6 as neutral, and 7-9 as difficult. For hand movements with the sketchpad, 13% of participants perceived the VH feedback condition to be difficult, and neither the VHA or the VA conditions were considered to be difficult. However, for the arm movement with the tangible handle on the wall-display, both the VH and VHA conditions were not considered to be difficult. The VH condition in this case showed a reduced percentage on both the 'easy' and 'difficult' levels. The VA condition, on the contrary, showed an increased ratio on these two levels.

Following the task difficulty assessment, the workload required of participants was evaluated using the NASA-TLX assessment. Participants were asked to consider how demanding the rhythmic sketch was with respect to the different augmented feedback conditions. The results of hand and arm interactions are plotted in figure 11 respectively. A repeated-measures ANOVA revealed that for hand tasks with the touch-pad, a statistically significant effect was seen between the four feedback conditions over the five measurements: performance, physical demand, mental demand, frustration and effort (Table 5). Bonferroni Post hoc tests showed that there was no statistically significant difference between the evaluation of the VHA, VH and VA conditions, but all the three feedback conditions had statistically significant better evaluation than the baseline condition.

For the arm tasks with the tangible handle, statistical significance was revealed [405] in all measurements except for mental demand.

*5.3.2. User experience results*

Following the usability test with the task difficulty and workload assessments, participants were asked to evaluate their interactive experiences with the different types of augmented feedback. The evaluation process was designed [410] based on the Reaction cards [61, 62], a user experience evaluation tool that contains 118 words describing emotional reactions towards interaction and products. The seven most task-relevant words from each of the negative weighted, neutral weighted and positive weighted words sections were chosen [62] by three user experience researchers. The order of the questions relating to each of the 21 [415] adjectives were randomized in the user experience questionnaire. Participants rated each feedback condition against the 21 reaction words according to the intensity scale from 0 (not at all) to 10 (very much). The results of the subjective evaluation can be seen in figure 12. The individual ratings of the visual feedback had the highest score in the negative section and lowest score in the positive [420] section; while the VHA feedback had the highest score in the positive section and lowest score in the negative section. The rating on VH and VA feedback was close to each other and set in-between the V and VHA conditions. In the hand task with the touch-pad, the VH condition was rated slightly higher than the VA condition in the positive section. However, the higher rating of VH was [425] not maintained in the arm task on the wall-display, except that the rating for the confidence measurement was still very high.

**6. Discussion**

*6.1. Discussion of task performance results*

Correlation analysis showed that VH augmented terminal feedback enable [430] participants to perform rhythmic task that maintained an equivalent level of temporal accuracy compared with VA and VHA feedback. The same results were observed on the two interfaces, supporting different motion scales, but the performance on rhythm 1 was systematically lower than rhythms 2, 3, 4 in both cases as shown in figure 6. This observation indicates three things.

First, augmented haptic feedback can provide an alternative feedback strategy to augmented auditory feedback without declined rhythmic motor performance. Secondly, for the rhythmic sketch task on a sketchpad, which supports relatively small scale and fast hand movements, augmented haptic feedback combined with VA feedback does not diminish task performance, though the perceptual load might be increased due to the presence of an additional feedback mode. Third, rhythm 1 may be more difficult for most participants to entrain than the other three rhythms.

For the rhythmic sketch with the tangible handle on the wall-display, which were performed as an arm-based task, the two bimodal conditions in rhythms 2 and 3 produced the most accurate rhythmic sketches with higher correlation coefficients. In comparison, the bimodal advantage was not manifested in the hand tasks on the sketchpad (Figure 6 and table 1). The bi-modal feedback advantage has also been found in previous study with large scale arm movements on a different type of task (citation blinded for review), though the feedback was provided concurrently rather than as terminal feedback. The observation in current study reflects that for rhythmic arm movements, a two-modality combination could be the optimized feedback strategy. The different level of difficulty between the hand-based task and the arm-based task might be a plausible reason to account for the improved performance seen using the bi-modal display. One the one hand, increased task fulfilment time with the arm-based task requires the rhythmic pattern to be retained longer in the working memory system, leading to increased cognitive load [18]; on the other hand, the increased requirement of range of motion for the arm movement necessitates more attention to motor control, leading to increased perceptual load on the participants [63].

[460] As a result, with the terminal or concurrent feedback, all-three modalities might increase demands on both cognitive and perceptual resources and so diminish task performance. In deed, earlier research demonstrated that as task complexity increase, feedback strategies that can alleviate cognitive load becomes preferable [38]. The hand-based task might be easier to perform than the arm
[465] task given the smaller motion range and shorter task completion time, therefore the two bimodal conditions has no observable impact on task performance.

Another interesting observation is that the overall performance of arm movement on the wall-display has higher temporal accuracy than hand movement on the sketchpad, meanwhile, the levels of accuracy of arm movements under dif-
[470] ferent feedback conditions were closer together compared with that of hand movement. This may result from reduced movement speed while doing the rhythmic sketch by arm. With increased task execution time, the brain can process more information in the perception-action loop [64], which leads to improved control of the rhythmic movements. Even though the overall difference
[475] in temporal accuracy between the two scales of motion exists, the modulation effect of the feedback strategies on interaction performance was consistent across two interfaces.

Summarising the results of the task performance assessments the correlation analysis revealed (a) the equivalent effect of feedback strategy of VH and VA,
[480] (b) the bi-modal feedback advantage with increase task difficulty, and (c) the difference of rhythmic performance with two scales of motion range. However, these observations were obtained base on post-interaction data, thus could not reflect the quality of the rhythmic movement that was executed during the interaction. Due to this unknown interaction property, potential correlations
[485] between the quality of the movement and the interactive task performance could not be identified. Hence kinematic analysis was applied to analyze the quality of the movement with different types of augmented feedback.

*6.2. Discussion on kinematic results*

In the DTW distance analysis, the warping distance of time series data indicates the similarity of rhythmic patterns of movements which may have been executed with unequal durations. Since it was calculated based on participants' speed profiles, the closer the warping distance between two speed profiles, the greater is the similarity of motion control during the interaction. First of all, comparing the results of the DTW analysis (Figure 6) and the correlation analysis (Figure 7), the behavior patterns of the execution process and of the task performance results were compatible with both of the interaction scales, e.g. the higher correlation coefficient associated with the lower warping distance.

Specifically, for the hand movement with sketchpad, the warping distance of VH, VA to VHA feedback condition were shorter than that of V to VHA condition, meanwhile, the correlation coefficient between VH,VA and VHA were higher than the V condition. For the arm movement with the tangible handle on the wall-display, a similar pattern was less salient but still can be observed in three out of the four rhythms. Second, the DTW analysis reflected the quality of movements during task execution which was not accessible through correlation analysis. From the results of the correlation coefficient test, the hand and arm movements with the rhythm patterns in the two interfaces varies in the range from 0.55 to 0.81 , which indicates a medium to good association between the sketched rhythm and sample rhythm. However, the DTW analysis revealed a overall shorter warping distance in the arm tasks with the tangible handle than with the hand tasks with the sketchpad, which indicates a better control of rhythmic movement in the arm tasks. Given the above two points, the DTW technique in this case could be a valid and complementary approach for the further analysis of augmented feedback effects.

Comparing the warping distance between the V, VA and VH conditions with the VHA condition, the rhythmic sketch by hand under the augmented VA condition has the smallest warping distance in the sketchpad interaction. For all the four rhythms, both the VH and the VA conditions produced a smaller warping distance than the base line condition. These facts indicate that for the relatively small and fast hand movements using the sketchpad, augmented feedback conditions involving audio induced the best concordance of rhythmic motor control, while tactile stimuli produced less concordant effects. For the interaction with the tangible handle on the wall-display, augmented feedback conditions involving audio produced the same modulation effect on two out of the four rhythms. However there was no salient difference across feedback groups with the other two rhythms. In general the rhythmic sketch performed with the arm over a larger distance had a similar level of motor control irrespective of whether VH or VA were employed as the feedback strategy.

Comparing the warping distance of the two interactive scales, the overall rhythmic arm movements do not have as big a feedback difference as the hand movements. This means that with the rhythmic sketch task, arm movements have less variance than the comparatively fast and small hand movements. One possible explanation following the previous discussion could be that more attention has been allocated to the motor control for arm movements, while hand movements require less attention but was performed in a less controlled manner. Another explanation might be that the augmented multimodal terminal feedback had less of a modulating effect on rhythmic movement control in arm tasks. To determine which of these explanations, or some other explanation, is more plausible requires further investigation using qualitative approaches.

As discussed in section 6.1, the results of correlation analysis showed a similar performance between VH, VA, and VHA feedback conditions, however, the DTW distance analysis showed that the underlying movement quality may vary. At this point, it is fair to say that DTW analysis of rhythmic movements can be a complementary technique to provide an explanation of the correlation analysis of the interaction outcomes, as well as bringing insight concerning the quality of motion control during the interaction. The last point may be of particular valuable in the areas of motor training and rehabilitation.

Another indicator used for the kinematic analysis was movement smoothness, which was determined by the number of velocity peaks during the execution of an action. In general, results showed that the VH feedback condition lead to the best movement smoothness with the least velocity peaks in both of the hand and arm tasks on the two interfaces. The facilitatory effect of VH feedback was more salient in the arm tasks than the hand tasks as shown in figure 8. However, comparison with the baseline V condition of VA and VHA condition showed inconsistent results on two interfaces. For the hand movement with sketchpad, the VA feedback lead to more velocity peaks than the base-line condition, and the VHA condition did not show improved smoothness compared to the baseline condition. For the arm movement with the tangible handle , improved smoothness of these two manipulation conditions has been observed.

Comparing this observation with the results from the correlation analysis, the VA and VHA feedback conditions in the hand tasks lead to similar or worse movement smoothness, e.g. produced similar or more velocity peaks during task execution, but have better task performance than the baseline condition, e.g. had higher correlation coefficient (Table 1). But in the arm tasks with the tangible handle, better movement smoothness and better task performance was produced under both the VA and VHA conditions. Given above two facts, it is reasonable to say that the quality of rhythmic sketch performance may not correlated with the smoothness of motion execution. The feedback strategy, the motion range, e.g. on the sketchpad or on the wall-display, and the time of execution, e.g. fast hand movements or slow arm movements, could both be factors that influence the smoothness of rhythmic movements.

Comparing the velocity peaks calculation with the DTW analysis of hand tasks, which showed that feedback condition with audio supported the best concordance of motor control with the shortest DTW distance, while the number of velocity peaks produced with VA feedback was highest. One plausible explanation could be that improved motor control leads to more velocity peaks during task execution. However, this explanation does not fully account for the facts observed in the arm tasks, that the warping distance of feedback conditions with audio produced the shortest DTW distance with increased number of velocity peaks in just 2 out of 4 rhythms. Feedback conditions with vibrotactile,

580 however, lead to a similar level of DTW distance to VA condition with reduced number of velocity peaks. This fact indicates that as the arm tasks have a relatively large motion range and reduced movement speed, vibrotactile in the multimodal feedback supports better movement smoothness without decreaed motor control capacity.

585 As an interim summary of the kinematic analyses based on two measurements, three outcomes need to be highlighted. First, combining the correlation analysis with DTW analysis, a high quality of motor control during task execution, e.g. a short warping distance, could indicate a good interaction performance, e.g. high correlation coefficient, but a good performance measure does

590 not indicate a high quality of motor control. Furthermore, combining the correlation analysis with velocity peak calculation, there was no correlation between interaction performance measures and the quality of motion smoothness. Second, the DTW analysis can be used to reveal the quality of motor control during the rhythmic sketch. It reflected that the rhythmic sketches in arm tasks were

595 systematically better than those in hand tasks, but the reason to account for this observation is not certain at this point. Another measurement, the velocity peaks revealed the movement smoothness between the four main movement phases. The feedback strategy, with this analysis, showed a larger effect in the arm tasks with the wall-display than hand tasks with the sketchpad. Third, the

600 VA and VH feedback conditions lead to improved motor control and movement smoothness compared with the baseline condition. While VA feedback had a larger facilitatory effect on improving movement concordance in the hand tasks and VH feedback on improving movement smoothness in both tasks.

*6.3. Discussion on qualitative observation*

605 The qualitative data were analyzed from two perspectives, the usability perspective and the user experience perspective. The first part of the discussion was focused on usability, which was analyzed based on subjective evaluation of task difficulty and NASA-TLX workload evaluation. The second part of the evaluation was focused on user experience, which was evaluated based on subjective [610] rating with Reaction cards [61, 62].

The subjective evaluation of task difficulty in relation to feedback strategy showed that the VHA feedback condition was considered to be the most easily performed task. The VH and VA feedback were rated harder than the VHA condition, but easier than the baseline condition. The same results were observed [615] with both interfaces. Meanwhile, the scores for each feedback condition were very close between the two scales of interfaces (Figure 10). This indicates that the perceived difficulty level of rhythmic sketching is not influenced, in these particular cases, very much by tasks with different scales of movement, though the task performance and quality of the movements do varies.

[620] The frequency of rating showed that for the hand task, all the ratings for VA and VHA feedback conditions did not fall into the 'difficult' level, while the VH had a small ratio of ratings (13%) that fell in the 'difficult' level (Figure 9). With respect to arm tasks, the VA condition had a ratio that fell in the 'difficult' level while the ratio of 'difficult' level for VH condition decrease to [625] 0. Meanwhile, the rating for VHA condition had an increased ratio in the 'easy' level. These observations reflected that feedback conditions that include vibrotactile stimulus could be preferable in the arm tasks.

NASA-TLX workload evaluation showed that the perceived workload for hand tasks do not have a statistically significant difference among VH, VA and [630] VHA conditions, but compared with the baseline condition, all these three manipulation conditions had significantly higher scores in relation to performance, and lower scores in relation to physical and mental demand, frustration and effort. This indicates that the VH augmented feedback for rhythmic sketching had equivalent effect on perceived workload with both the VA and VHA [635] conditions.

For the arm tasks, perceived mental demand was high in both the manipulation conditions and the baseline condition without a statistically significant difference being observed. Comparing with hand tasks, this increased mental demand (as well as increased physical demand) on arm tasks could explain one

640 of the results obtained in DTW analysis, that warping distance in arm movements was systematically lower than that in hand tasks. The increased physical and mental demands during interaction could account for the increased need for attention to motor control, as discussed in previous section. Thus the rhythmic sketches produced by arm have systematically shorter DTW distance than those

645 produced by hand movements. Regarding the other evaluation measures for arm tasks in the NASA-TLX assessment, a statistical significance was only observed in the feedback conditions with vibrotactile stimuli, which has lower physical demand, frustration and effort than the baseline condition. This result reflected that for the arm task, the feedback with rhythmic vibrotactile information was

650 able to facilitate arm motion, reduce task frustration and reduce the over all effort.

From the user experience evaluation perspective, the results of the subjective rating showed a consistency across negative, neutral and positive categories (Figure 12). The rating of the baseline condition (V feedback) was highest in the

655 negative category and lowest in the positive category, while the VHA feedback was rated the other way around. The rating of the VH and VA conditions were close to each other except on the measure of confidence, and the ratings came in-between the baseline condition and the VHA condition for the hand tasks with sketchpad. For the arm tasks, the rating of 'predictable', 'controllable'

660 and 'confidence' for VH feedback in the positive category was increased and higher than the VA feedback. This indicates that the augmented feedback with vibrotactile information increases the level of confidence. The ratings of the level of confidence also showed higher scores in the arm tasks than the hand tasks. Combining these results with the smoothness analysis, which showed that the VH condition for the arm tasks lead to the least velocity peaks, we could deduce that participants' confidence level in their task execution is correlated with their movement smoothness, and the number of velocity peaks of movements could be an indicator of measuring user's confidence level of rhythm-related interactive tasks.

As an interim summary of the discussion on qualitative analysis, the single ease-of-use evaluation, NASA-TLX workload evaluation and the user experience evaluation provide complimentary results from different perspectives. The single ease-of-use evaluation revealed more preference for feedback strategies involving vibrotactile stimuli in the arm tasks with larger motion range. The NASA-TLX assessment showed the equivalent effect between VH and VA feedback on perceived workload for hand tasks, as well as increased mental demand in arm tasks across all feedback strategies. It is also in agreement with the DTW analysis, in that high mental and physical demand correlated to low warping distance between speed profiles. Last but not least, the user experience evaluation revealed that feedback with vibrotactile stimuli had the effect of improving interaction confidence and assurance, which was consistent with the objective evaluation on smoothness analysis. Given this last point, we propose that smoothness analysis could be a new indicator of confidence level in motion-relevant applications.

## 7. Conclusion

This paper presents two studies with two research objectives. First, we observed and evaluated whether providing haptic feedback have equivalent facilitatory effects compared to audio feedback on people's rhythmic movement when combined with visual information; and whether similar behavioral patterns can be observed on different scales of motion range on two interfaces; second, we introduced kinematic analysis as a complementary evaluation method to reveal hidden interactive patterns during rhythmic sketching with augmented feedback. Three analysis methods were applied in the two studies: the task performance analysis, kinematic analysis and qualitative analysis. In summary, our results suggest the following:

First, VH augmented feedback for rhythmic sketch tasks showed equivalent facilitatory effect with VA and VHA feedback in terms of temporal accuracy. This result is obtained from both of the interfaces with two scales of sketch respectively. Though the task performance may also vary with the participants ability to follow the rhythm.

Second, for the purpose of achieving better interaction performance, the bi-modal feedback strategy is more effective than the tri-modal strategy for relatively slow movements and the large scale motion range, since the tri-modal strategy may cause perceptual overload. For the purpose of improving precision of rhythmic movements or gestures, feedback involving auditory stimuli can be more effective. While for supporting better movement smoothness, feedback with vibrotactile stimuli is preferable.

Third, the DTW analysis showed consistent results with task performance analysis on the one hand, and revealed underlying variance on motor control ability on the other hand. The smoothness analysis based on the number of velocity peaks was related to the high subjective rating on interaction confidence. We broadly proposed that it could be a new behavioral measure for decoding and accessing emotional state during sketch-related interaction tasks.

Last but not least, the combination of task performance analysis, kinematic analysis and subjective evaluation can compliment each other and leads to better tools to evaluate and understand the mechanisms underlying the rhythmic motion with multimodal perception.

Given the fact that VH feedback showed equivalent effects on task performance in terms of the temporal accuracy, future studies could investigate the optimization of this feedback strategy for supporting better rhythmic motor control or training. Future studies could also adopt the same feedback strategy for investigations on more generalised gesture or sketch-related tasks. Based on the approach of obtaining the DTW distance and movement smoothness during the interaction, the two streams of information could be sonified or visualized in real-time. Therefore, further study could also investigate how to design the real-time display for improving motivation of motor training or self-directed rehabilitation at home. One limitation of the current study is the small sample size. Though the experimental conditions were strictly controlled, these results still need to be interpreted with caution. Further investigation and verification is also required.

[8] A. O. Effenberg, U. Fehse, G. Schmitz, B. Krueger, H. Mechling, Movement sonification: effects on motor learning beyond rhythmic adjustments, Frontiers in neuroscience 10 (2016) 219.

[9] A. Singh, S. Piana, D. Pollarolo, G. Volpe, G. Varni, A. Tajadura-Jiménez, A. C. Williams, A. Camurri, N. Bianchi-Berthouze, Go-with-the-flow: tracking, analysis and sonification of movement and breathing to build confidence in activity despite chronic pain, Human–Computer Interaction 31 (3-4) (2016) 335–383.

[10] S. Schneider, P. W. Schönle, E. Altenmüller, T. F. Münte, Using musical instruments to improve motor skill recovery following a stroke, Journal of neurology 254 (10) (2007) 1339–1346.

[11] A. Bouwer, S. Holland, M. Dalgleish, The haptic bracelets: learning multilimb rhythm skills from haptic stimuli while reading, in: Music and human-computer interaction, Springer, 2013, pp. 101–122.

[12] B. H. Repp, Y.-H. Su, Sensorimotor synchronization: a review of recent research (2006–2012), Psychonomic bulletin & review 20 (3) (2013) 403–452.

[13] M. Turk, Multimodal interaction: A review, Pattern Recognition Letters 36 (2014) 189–195.

[14] E. Frid, J. Moll, R. Bresin, E.-L. S. Pysander, Haptic feedback combined with movement sonification using a friction sound improves task performance in a virtual throwing task, Journal on Multimodal User Interfaces (2018) 1–12.

[15] J. Rodríguez, T. Gutiérrez, O. Portillo, E. J. Sánchez, Learning force patterns with a multimodal system using contextual cues, International Journal of Human-Computer Studies 110 (2018) 86–94.

Table 3: Number of velocity peaks in four main movement phases

| Interfaces | Rhythms | | VHA | VA | VH | V |
|---|---|---|---|---|---|---|
| Sketchpad | 1 | $\overline{X}$ | 10.73 | 11.96 | 10.19 | 9.67 |
| | | $\sigma$ | 4.39 | 4.08 | 4.68 | 4.71 |
| | 2 | $\overline{X}$ | 10.94 | 11.63 | 9.54 | 10.63 |
| | | $\sigma$ | 4.17 | 3.80 | 4.12 | 4.88 |
| | 3 | $\overline{X}$ | 9.75 | 12.15 | 10.35 | 11.02 |
| | | $\sigma$ | 4.09 | 4.33 | 4.34 | 4.34 |
| | 4 | $\overline{X}$ | 10.17 | 11.44 | 9.75 | 10.38 |
| | | $\sigma$ | 3.45 | 3.50 | 3.68 | 3.91 |
| All rhythms | | $\overline{X}$ | 10.40 | 11.79 | 9.96 | 10.42 |
| | | $\sigma$ | 4.07 | 3.95 | 4.23 | 4.45 |
| Wall-display | 1 | $\overline{X}$ | 11.65 | 11.81 | 10.17 | 10.04 |
| | | $\sigma$ | 6.86 | 5.39 | 5.36 | 5.46 |
| | 2 | $\overline{X}$ | 13.94 | 11.46 | 8.90 | 10.63 |
| | | $\sigma$ | 9.55 | 5.93 | 4.78 | 6.07 |
| | 3 | $\overline{X}$ | 13.00 | 11.17 | 9.21 | 11.02 |
| | | $\sigma$ | 7.86 | 5.59 | 4.42 | 6.25 |
| | 4 | $\overline{X}$ | 14.60 | 12.06 | 10.35 | 11.19 |
| | | $\sigma$ | 9.58 | 6.69 | 6.10 | 7.02 |
| All rhythms | | $\overline{X}$ | 13.30 | 11.62 | 9.66 | 10.72 |
| | | $\sigma$ | 8.61 | 5.93 | 5.24 | 6.24 |

Table 4: Wilcoxon Signed-Ranks test results

| Ranks | VHA-VH | VHA-VA | VHA-V | VH-VA | VH-V | VA-V |
|---|---|---|---|---|---|---|
| Sig.(hand) | .008 | .016 | .001 | .096 | .001 | .001 |
| Sig.(arm) | .021 | .176 | .001 | .163 | .001 | .000 |

Table 5: Statistical results of Repeated-measures ANOVA for NASA-TLX workload evaluation.

|  | Measurements | $F(3, 13)$ | p | $\eta^2$ | Bonferroni test |
|---|---|---|---|---|---|
| Hand tasks | Performance | 13.946 | .000 | .482 | VHA (72.188) to V (47.188), $p = .000$, VH (71.688) to V, $p = .000$, VA (70.125) to V, $p = .016$ |
|  | Physical demand | 8.614 | .000 | .365 | VHA (26.188) to V (47.688), $p = .013$, VH (33.000) to V, $p = .042$, VA (34.063) to V, $p = .012$ |
|  | Mental demand | 13.396 | .000 | .472 | VHA (23.813) to V (47.875), $p = .000$, VH (30.438) to V, $p = .024$, VA (32.063) to V, $p = .004$ |
|  | Frustration | 46.578 | .000 | .756 | VHA (26.625) to V (29.250), $p = .000$, VH (30.063) to V, $p = .000$, VA (64.375) to V, $p = .000$ |
|  | Effort | 11.582 | .000 | .436 | VHA (47.938) to V (71.563), $p = .000$, VH (50.563) to V, $p = .000$, VA (51.313) to V, $p = .000$ |
| Arm tasks | Performance | 3.120 | .035 | .172 | VHA (68.625), VH (67.938), VA (72.563), V (58.438) |
|  | Physical demand | 7.627 | .000 | .337 | VHA (34.750) to V (52.250), $p = .002$, VH (31.125) to V, $p = .025$, VA (53.063) |
|  | Mental demand | 1.213 | .637 | .015 | VHA (65.500), VH (65.313), VA (57.188), V (65.625) |
|  | Frustration | 15.357 | .000 | .506 | VHA (28.000) to V (60.125), $p = .000$, VH (32.813) to V, $p = .000$, VA (40.751) to V, $p = .028$ |
|  | Effort | 4.683 | .006 | .238 | VHA (57.500), VH (58.125) to V(78.750), $p = .007$, VA (63.938) |